\documentclass[%
 aip,
 jmp,%
 amsmath,amssymb,
 reprint,%
]{revtex4-1}

\usepackage{graphicx}
\usepackage{dcolumn}
\usepackage{bm}
\usepackage{natbib}

\newcommand\curl{\text{\rm curl\,}}

\begin{document}


\title{A generalization of vortex lines}

\author{M. Fecko}
\email{fecko@fmph.uniba.sk}
\affiliation{Department of Theoretical Physics, Comenius University in Bratislava, Slovakia}



\begin{abstract}
Helmholtz theorem states that, in ideal fluid, vortex lines move with the fluid.
Another Helmholtz theorem adds that strength of a vortex tube is constant along the tube.
The lines may be regarded as integral surfaces of a 1-dimensional integrable distribution
(given by the vorticity 2-form).
In general setting of theory of integral invariants, due to Poincar\'e and Cartan,
one can find $d$-dimensional integrable distribution  (given by a possibly higher-rank form)
whose integral surfaces show both properties of vortex lines:
they move with (abstract) fluid and, for appropriate generalization of vortex tube,
strength of the latter is constant along the tube.
\end{abstract}

\maketitle

\tableofcontents

\section{\label{sec:intro}Introduction}

In hydrodynamics, \emph{vortex lines} are field lines of {\it vor\-ti\-ci\-ty} vector field
$\boldsymbol \omega$,
which is curl of velocity field $\mathbf v$.
\emph{Vortex tube} is a surface made 
of vortex lines passing through each point of a transversal circuit
(so that the circuit then encircles the tube).

It was as early as 1858 (see Ref.~\onlinecite{Helmholtz1858} and
     Refs.~\onlinecite{Truesdell1954, Saffman1992, Batchelor2002, WuMaZhou2006, ThorneBlandford2017})
that Helmholtz proved that, in the case of ideal
and barotropic fluid that is only subject to conservative forces,

- vortex lines ``move with the fluid''
  (the same fact is sometimes expressed as that the lines are ``frozen into the fluid"
   or that ``vortex lines are material lines'') and that

 - strength of a vortex tube is the same at all cross-sections.

Here the strength is defined
as the flux of the vorticity field $\boldsymbol \omega$ for cross-section itself
or, via Stokes theorem,
as circulation of the velocity field $\mathbf v$ round the circuit cut by the cross-section.

\emph{Geometrical} (and even topological) language has proved very effective in hydrodynamics since a long time ago.
In particular, for obtaining and classifying of \emph{conserved quantities},
one can use \emph{Hamiltonian} structure of hydrodynamic equations
or the interconnection of \emph{symmetries} and conserved quantities,
see e.g. Refs. ~\onlinecite{ArnoldKhesin1998, Anco2013, AncoDarTufail2015, BesseFrisch2017}.

When treated geometrically, the Helmholtz statements may get specific meaning.

For example, Arnold succeeded to show (see Ref.~\onlinecite{Arnold1966}),
that the Euler equation for incompressible fluid on $n$-dimensional Riemannian manifold
has an elegant formulation as the geodesic equation on the Lie group
of volume-preserving diffeomorphisms of the given manifold.
(In strong analogy with a much simpler - finite-dimensional - description of a rotating top,
 where the Lie group is $SO(3)$.)
In this approach, Helmholtz theorem stems from
invariance of \emph{coadjoint orbits} with respect to the dynamics.
(For subsequent work in this direction,
 see Refs. ~\onlinecite{GuilleminSternberg1980, MarsdenWeinstein1983, MarsdenRatiuWeinstein1984, Novikov1982, KhesinChekanov1989},
 and, in particular, monography ~\onlinecite{ArnoldKhesin1998}).

The point of view this paper is based on starts from regarding hydrodynamics of ideal fluid
as an application of the theory of \emph{integral invariants} due to Poincar\'e and Cartan
(see Refs.~\onlinecite{Poincare1899, Cartan1922},
     Ref.~\onlinecite{Gantmacher1975}            or, in mo\-dern presentation,
     Ref.~\onlinecite{Arnold1989, LibermannMarle1987, Kiehn1975, Fecko2013}).
Then, original Poincar\'e version of the theory refers to \emph{stationary} (time-independent) flow,
described by stationary Euler equation,
whereas Cartan's extension embodies the full, possibly time-dependent, situ\-ation.

Let us remark that although integral invariants due to Poincar\'e and Cartan
are mostly known from classical Hamiltonian mechanics,
see e.g. Ref.~\onlinecite{LandauLifshitz1995}, its realm of applications is wider
(see Refs.~\onlinecite{Cartan1922, Gantmacher1975}).

The idea of a proof of Helmholtz theorem on vortex lines might go,
within the integral invariants setting, as follows (for details, see below).
First, vortex lines are identified with integral surfaces of an 1-di\-men\-sional
integrable distribution, defined in terms of an ap\-pro\-pri\-ate 2-form.
Second, structure of the (Euler) equation of motion immediately reveals
that the 2-form is \emph{Lie-invariant} w.r.t. the flow of the fluid.
So, third, the corresponding distribution is invariant w.r.t. the flow and, consequently,
its integral surfaces are invariant w.r.t. the flow of the fluid.
But this is exactly what Helmholtz statement says.

Now, it turns out that the same reasoning may be repeated
within the \emph{general integral invariant} setting
(so beyond even the ``$n$-dimensional Riemannian hydrodynamics'').
What differs is that we have an integrable distribution
based on a possibly \emph{higher-degree} Lie-invariant differential form, there.
In particular, the distribution may be \emph{higher-dimensional} and, consequently,
its integral surfaces become then higher-di\-men\-sional, too.
Never\-theless, they still obey the Helmholtz-like rule of ``moving with the fluid''
(i.e. the \emph{abstract} flow in the general theory
 translates the in\-te\-gral surfaces into one another).

Concerning the vortex \emph{tubes} Helmholtz theorem,
proof of the original statement is very easy and corresponding
generalization to integral invariants setting is almost self-evident.

The structure of the paper is as follows.

In Section \ref{subsec:poincare}, in order to make the text self-contained,
we shortly remind the reader, in modern language,
of the \emph{Poincar\'e} theory of integral invariants.
Then, in Section \ref{subsec:eulertimeindependent},
we present \emph{stationary} Euler equation rewritten in a form
needed for profiting from the Poincar\'e theory.
Sections \ref{subsec:helmholtztimeindependent} and \ref{subsec:helmholtztubestimeindependent}
then show how (easily)
one obtains Helmholtz results within this scheme.

The same program is then repeated, for the case of time-\emph{dependent} Euler equation
(based on \emph{Cartan}'s extension of the theory of integral invariants), in Sections
\ref{subsec:cartan},
\ref{subsec:eulertimedependent},
\ref{subsec:helmholtztimedependent} and
\ref{subsec:helmholtztubestimedependent} .

Finally, as the principal topic of the paper, general, possibly higher-dimensional
\emph{surfaces} moving with the (abstract) fluid in the phase space of a system,
are studied in Sections 
\ref{subsec:surfacespoincare} (stationary case) and
\ref{subsec:surfacescartan}
(time-dependent case; here also \emph{extended} phase space plays a role).

\section{\label{sec:timeindependent}Time-independent flow}

\subsection{\label{subsec:poincare}Poincar\'e integral invariants}

Consider a manifold $M$ endowed with dynamics given by a \emph{vector field} $v$
\begin{equation} \label{dynamicsgivenbyv}
       \dot \gamma = v
       \hskip 1.5cm
       {\dot x}^i = v^i(x)
\end{equation}
The field $v$ generates the dynamics (time evolution) via its flow $\Phi_t \leftrightarrow v$.
We will call the structure \emph{phase space}
\begin{equation} \label{defphasespace}
      (M,\Phi_t \leftrightarrow v) \hskip 1cm \text{\emph{phase space}}
\end{equation}
In this situation, let us have a $k$-form $\alpha$ and consider its integrals over various
$k$-chains ($k$-dimensional surfaces) $c$ on $M$.
Due to the flow $\Phi_t$ corresponding to $v$, the $k$-chains flow away, $c \mapsto \Phi_t (c)$.
Compare the value of integral of $\alpha$ over the original $c$ and integral over $\Phi_t (c)$.
If, {\it for any chain} $c$, the two integrals are equal, it reflects a remarkable
property of the form $\alpha$ with respect to the field $v$. We call it integral invariant:
\begin{equation} \label{integrinvariant}
       \int_{\Phi_t (c)} \alpha = \int_c \alpha
       \hskip .5cm \Leftrightarrow \hskip .5cm
       \int_c \alpha \  \  \  \text{is \emph{integral invariant}}
\end{equation}
For {\it infinitesimal} $t\equiv \epsilon$ we have
\begin{equation}
      \label{odtec1}
       \int_{\Phi_\epsilon (c)} \alpha
       = \int_c \alpha + \epsilon \int_c \mathcal L_v\alpha
\end{equation}
(plus, of course, higher order terms in $\epsilon$; here $\mathcal L_v$ is \emph{Lie derivative} along $v$).
Since (\ref{integrinvariant}) is to be true {\it for each} $c$, we get from (\ref{odtec1})
\begin{equation} \label{podmienka2}
       \mathcal L_v\alpha = 0
\end{equation}
This is the \emph{differential version} of the statement (\ref{integrinvariant}).

In specific situations, it may be enough that some integral only behaves invariantly when restricted to
an important sub-class of $k$-chains, namely $k$-\emph{cycles}.
These are chains whose boundary vanish:
\begin{equation} \label{defcyklu}
    \partial c = 0 \hskip 2cm c = \ \text{\emph{cycle}}
\end{equation}
If this is the case, the condition (\ref{podmienka2}) is overly strong. It can be weakened to
\begin{equation} \label{podmienka3}
       \mathcal L_v\alpha = d\tilde \beta
\end{equation}
for some form $\tilde \beta$.
(The form $\mathcal L_v\alpha$ may just be \emph{exact} rather than vanish.)
Indeed, in one direction, Eqs. (\ref{defcyklu}) and (\ref{podmienka3}) then give
\begin{equation} \label{podmienka33}
       \int_c \mathcal L_v\alpha  = \int_c d\tilde \beta = \int_{\partial c} \tilde \beta = 0
\end{equation}
so that (\ref{integrinvariant}) \emph{is} fulfilled.
In the opposite direction, if (\ref{integrinvariant})  is to be true for \emph{each cycle},
the form under the last integral sign in (\ref{odtec1}) is to be \emph{exact}
due to \emph{de Rham theorem}, so (\ref{podmienka3}) holds.

According to whether the integrals of forms are in\-vari\-ant for arbitrary $k$-chains or just for $k$-cycles,
integral invariants are known as either \emph{absolute} invariants (for any $k$-chain) or \emph{relative} ones
(just for $k$-cycles;
 notice that $\mathcal L_v(d\alpha) = 0$ holds from (\ref{podmienka3}),
 so whenever $\alpha$ gives relative invariant,
 $d\alpha$ already gives an \emph{absolute} one).

Now, let us see what we can say about \emph{relative} integral invariants.
The condition (\ref{podmienka3}) may be rewritten,
using Cartan's formula
\begin{equation} \label{cartanmagic}
                 i_vd+di_v = \mathcal L_v
\end{equation}
as
\begin{equation} \label{ivdalphajeexaktna}
        i_vd\alpha = d\beta
\end{equation}
(where $\beta = \tilde \beta - i_v\alpha$).
Therefore the following main statement on relative invariants is true:
\begin{equation} \label{jetotoiste2}
        i_vd\alpha = d\beta
        \hskip .5cm \Leftrightarrow \hskip .5cm
        \oint_c\alpha = \ \ \text{\emph{relative} invariant}
\end{equation}
So we can identify the presence of relative integral invariant \emph{in differential version}:
on phase space $(M,v)$, we find a form $\alpha$
such that the l.h.s. of Eq. (\ref{ivdalphajeexaktna}) is exact.

\subsection{\label{subsec:eulertimeindependent}Stationary Euler equation}

\emph{Euler equation} for ideal (inviscid) fluid
\begin{equation} \label{eulernorm}
                        \rho \left(\partial_t \bold v + (\bold v \cdot \boldsymbol \nabla) \bold v \right)
                        =
                        - \boldsymbol \nabla p -\rho \boldsymbol \nabla \Phi
\end{equation}
(see, e.g. Ref.~\onlinecite{LandauLifshitz1987, Batchelor2002})
reduces, for \emph{stationary} flow, to
\begin{equation} \label{eulerstatnorm1}
                      (\bold v \cdot \boldsymbol \nabla) \bold v
                   =
                   - \frac{1}{\rho} \boldsymbol \nabla p - \boldsymbol \nabla \Phi
\end{equation}
Here mass density $\rho$, velocity field $\bold v$, pressure $p$ and potential
$\Phi$ of the volume force field are
functions of $\bold r$. 

In general, equation of state of the fluid may be written as
\begin{equation} \label{generalfluid}
                 p=p(\rho, s) \hskip 1cm \text{general fluid}
\end{equation}
where $s$ is (specific) entropy (i.e. entropy per unit mass).
However, one can think of an important model, where the pressure depends \emph{on} $\rho$ \emph{alone}:
\begin{equation} \label{barotropicfluid}
                 p=p(\rho)
                 \hskip 1.3cm \text{\emph{barotropic} fluid}
\end{equation}
In this case, there exists $P(\bold r)$, called specific \emph{enthalpy}, such that
\begin{equation} \label{defP}
                 \frac{1}{\rho} \boldsymbol \nabla p = \boldsymbol \nabla P
\end{equation}
and (\ref{eulerstatnorm1}) takes the form
\begin{equation} \label{eulerstatnorm2}
                      (\bold v \cdot \boldsymbol \nabla) \bold v
                   =
                   - \boldsymbol \nabla (P + \Phi)
\end{equation}
Now it turns out (check in Cartesian coordinates)
that Eq. (\ref{eulerstatnorm2}) may be rewritten in the form of Eq. (\ref{ivdalphajeexaktna})
for the particular choice $\alpha = \tilde v$ and $\beta = -\mathcal E$, i.e. as
\begin{equation} \label{eulerstatform3}
                 i_v d\tilde v = - d\mathcal E
                 \hskip 1cm
                 \text{\emph{Euler equation}}
\end{equation}
(stationary and barotropic), where
\begin{equation} \label{deftildev}
                      \tilde v := \bold v \cdot d\bold r \hskip .5cm (\equiv \flat_g v \ \equiv g(v, \ \cdot \ ))
\end{equation}
is the covector (= 1-form) associated with the velocity vector field $v = v^i\partial_i$
 in terms of ``lowering of index'' ($\equiv \flat_g$ procedure) and
 \begin{equation} \label{deffunctionE}
                 \mathcal E := v^2/2 +P+ \Phi
                  \hskip 1cm
                 \text{\emph{Bernoulli function}}
\end{equation}
The \emph{vorticity 2-form} $d\tilde v$, present in Eq. (\ref{eulerstatform3}),
is of crucial importance for us. We have
\begin{eqnarray} 
      \label{velocityform}
            \tilde v  &=& \bold v \cdot d\bold r \\
      \label{vorticityform}
           d\tilde v  &=& (\curl \bold v) \cdot d\bold S \ \equiv \ \boldsymbol \omega \cdot d\bold S \\
      \label{vortexlinesexpr}
           i_{\gamma'}d\tilde v  &=& (\boldsymbol \omega \times {\bold r}') \cdot d\bold r
\end{eqnarray}
(see , e.g. \$8.5 in Ref.~\onlinecite{Fecko2006}) so that,
first, $d\tilde v$ indeed encodes complete information
about vorticity vector field $\boldsymbol \omega$
and, second, the equation
 \begin{equation} \label{vortexlinesexpr2}
                 i_{\gamma'}d\tilde v =0
                 \hskip 1cm
                 \text{\emph{vortex line equation}}
\end{equation}
expresses the fact that $\gamma (\lambda) \leftrightarrow \bold r(\lambda)$
corresponds to vortex line (the prime symbolizes tangent vector w.r.t. parameter $\lambda$;
particular parametrization is, however, irrelevant).

The form (\ref{eulerstatform3}) of the Euler equation turns out to be very convenient.
Short illustration:

1. Application of $i_v$ on both sides gives
\begin{equation} \label{bernoulli1}
                 v\mathcal E =0
                 \hskip 1cm
                 \text{\emph{Bernoulli equation}}
\end{equation}
(saying that $\mathcal E$ is constant along \emph{stream}lines).

2. Application of $i_{\gamma'}$ on both sides (where $\gamma'$ is from (\ref{vortexlinesexpr2})) gives
\begin{equation} \label{bernoulli11}
                 \gamma'\mathcal E =0
\end{equation}
(saying that $\mathcal E$ is constant along \emph{vortex}-lines).

3. Putting $d\tilde v =0$ (\emph{irrotational} flow) leads to
\begin{equation} \label{bernoulli2}
                 \mathcal E = \ \text{const.}
\end{equation}
(a version of Bernoulli equation saying that $\mathcal E$ is, then, constant throughout the fluid).

4. Just looking at (\ref{jetotoiste2}), (\ref{eulerstatform3}) and (\ref{deftildev}) one obtains
\begin{equation} \label{Kelvin_stat1}
                 \oint_c \bold v \cdot d\bold r = \text{const.}
                 \hskip 1.5cm \text{\emph{Kelvin's theorem}}
\end{equation}
(velocity circulation is conserved quantity).

5. Application of $d$ on both sides gives \emph{Helmholtz theorem}
(see the next Section \ref{subsec:helmholtztimeindependent}).

\subsection{\label{subsec:helmholtztimeindependent}Helmholtz statement on vortex lines - stationary case}

Application of $d$ on both sides of (\ref{eulerstatform3}) and using
(\ref{cartanmagic}) results in
\begin{equation} \label{vorticityisinv1}
                 \mathcal L_v (d\tilde v) = 0
\end{equation}
This is, however, nothing but infinitesimal version of the statement
\begin{equation} \label{vorticityisinv2}
                 \Phi_t^* (d\tilde v) = d\tilde v
                 \hskip 2cm
                 \Phi_t \leftrightarrow v
\end{equation}
or, in words, that the vorticity 2-form $d\tilde v$ is \emph{invariant}
w.r.t. the flow of the fluid.

Now, we can define a \emph{distribution} $\mathcal D$ in terms of $d\tilde v$:
\begin{equation} \label{distributiondef}
         \mathcal D
         :=
         \{ \text{vectors} \  w  \ \text{such that} \ \  i_w d\tilde v = 0 \ \ \text{holds}  \}
\end{equation}
Due to Frobenius criterion the distribution is integrable. Indeed, let
$w_1,w_2\in \mathcal D$. Then, because of the identity
\begin{equation} \label{identity1}
         i_{[w_1,w_2]} = [\mathcal L_{w_1}, i_{w_2}]
                  \equiv \mathcal L_{w_1} i_{w_2} - i_{w_2} \mathcal L_{w_1}
\end{equation}
(see, e.g., Ch.5.Ex.21 in Ref.~\onlinecite{CrampinPirani1986} or \$6.2 in Ref.~\onlinecite{Fecko2006})
plus (\ref{cartanmagic})
one immediately sees that
\begin{equation} \label{Disintegrable}
         i_{[w_1,w_2]} d\tilde v = 0
\end{equation}
i.e. $[w_1,w_2]\in \mathcal D$, too. So $\mathcal D$ is integrable.

From (\ref{vortexlinesexpr}) and (\ref{vortexlinesexpr2}) we see that the distribution
is 1-dimensional (in those points where $\boldsymbol \omega \neq 0$)
and that its integral surfaces are exactly vortex lines.
But this means that Helmholtz statement is true:
because of (\ref{vorticityisinv2}) and (\ref{distributiondef})
the distribution $\mathcal D$ is invariant w.r.t. $\Phi_t \leftrightarrow v$
and, consequently, its integral surfaces (i.e. vortex lines)
are invariant w.r.t. $\Phi_t \leftrightarrow v$, too.

\subsection{\label{subsec:helmholtztubestimeindependent}Helmholtz statement on vortex tubes - stationary case}

This statement is purely kinematical, it concerns the concept of vorticity itself.
It holds for \emph{arbitrary} velocity fields $v$,
even those which do not satisfy equations of motion
(so they cannot occur).

\begin{figure}[tb]
\begin{center}
\includegraphics[scale=0.30]{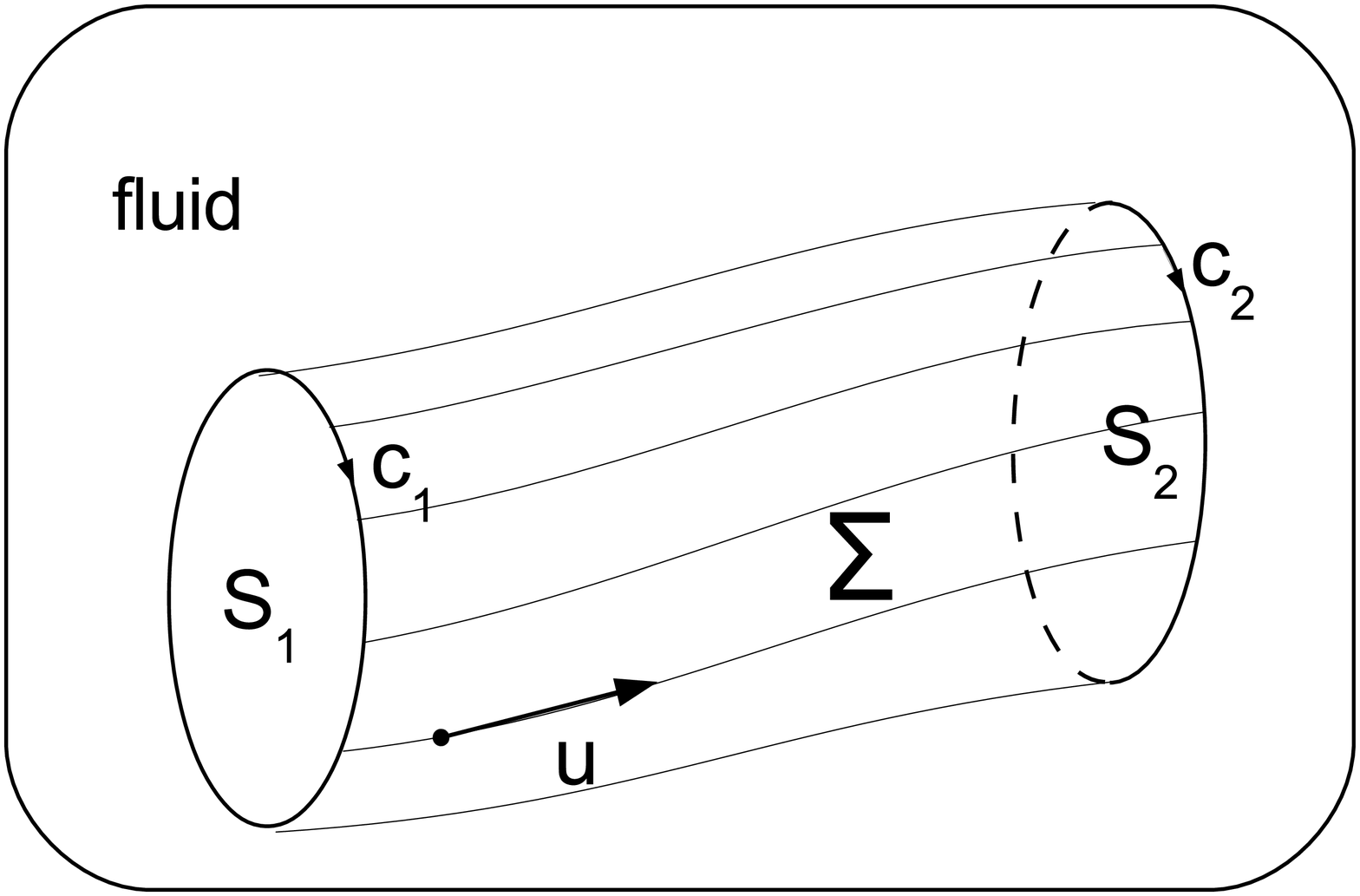}
\caption{Vortex tube $\Sigma$ is made of vortex lines emanating from (all points of) circuit $c_1 = \partial S_1$
         and entering the circuit $c_2 = \partial S_2$.
         Equation (\ref{jeabsolutny}) says that strength (vorticity flux) for the cross-section $S_1$
         is the same as the strength for the cross-section $S_2$.}
\label{helmholtztube}
\end{center}
\end{figure}
Let $u$ be a vector \emph{field} defined by $i_ud\tilde v = 0$, i.e.
a field tangent, at each point, to the vortex line passing through the point
(see Eq. (\ref{vortexlinesexpr2})).
Notice that any vortex line may be created from its single point by the flow $\Phi_s$ of $u$
and the same holds
(using evident freedom $u\mapsto fu$, $f$ being a function)
for the vortex \emph{tube} bounded by fixed circuits $c_1$ and $c_2$
(boundaries of fixed cross-sections $S_1$ and $S_2$, see Fig. \ref{helmholtztube}).

Consider the artificial (!) ``dynamics'' given by $u$. Then the equation $i_ud\tilde v = 0$
may be regarded as a particular case of the basic equation (\ref{jetotoiste2})
from the general theory of Poincar\'e integral invariants
(with $v\mapsto u$, $\alpha \mapsto \tilde v$ and $\beta \mapsto 0$).
So,
\begin{equation} \label{jerelativny}
                 \oint_c \tilde v \equiv \oint_c \bold v \cdot d\bold r \hskip .5cm \text{is relative invariant}
\end{equation}
and, consequently,
\begin{equation} \label{jeabsolutny}
                 \int_S d\tilde v \equiv \oint_S \boldsymbol \omega \cdot d\bold S \hskip .5cm \text{is absolute invariant,}
\end{equation}
both of them w.r.t. our ``\emph{artificial} dynamics'' generated by $u$
(as opposed to the \emph{real dynamics} generated by the fluid velocity field $v$).
Then, however, Eq. (\ref{jeabsolutny}) exactly says that the vorticity flux
does not depend on particular choice of cross-section $S$ cutting the tube.

Alternatively, one can use the proof of Eq. (\ref{jetotoiste4})
given in Appendix \ref{appproofintinv}
(with $\xi \mapsto u$ and $\sigma \mapsto \tilde v$).

\section{\label{sec:timedependent}Time-dependent flow}

\subsection{\label{subsec:cartan}Cartan integral invariants}

Cartan proposed, as a first step, to study the dynamics given in
(\ref{dynamicsgivenbyv}) and (\ref{defphasespace})
on $M \times \mathbb R$ (\emph{extended} phase space; \emph{time} coordinate is added)
rather than on $M$. Using the natural projection
\begin{equation} \label{projectiononm}
       \pi:M \times \Bbb R \to M
       \hskip .4cm (m,t)\mapsto m
       \hskip .4cm (x^i,t)\mapsto x^i
\end{equation}
the forms $\alpha$ and $\beta$ (from Poincar\'e theory) may be pulled-back from $M$ onto
$M \times \mathbb R$ and then combined into a single $k$-form
\begin{equation} \label{definiciasigma1}
       \sigma = \hat \alpha + dt \wedge \hat \beta
\end{equation}
(Here, we denote $\hat \alpha = \pi^*\alpha$ and $\hat \beta = \pi^*\beta$).
In a similar way, define a vector field
\begin{equation} \label{definiciaxi1}
      \xi = \partial_t +v
\end{equation}
Its flow clearly consists of the flow $\Phi_t \leftrightarrow v$ on the $M$ factor
combined with the trivial lapsing of time in the $\mathbb R$ factor.
Now a simple check (for which Appendix \ref{appdecomp} might come in handy) reveals that the equation
\begin{equation} \label{zakladnarovnica1}
      i_\xi d\sigma = 0
\end{equation}
is equivalent to (\ref{ivdalphajeexaktna}). And the main statement (\ref{jetotoiste2})
takes the form
\begin{equation} \label{jetotoiste4}
        i_\xi d\sigma = 0
        \hskip .5cm \Leftrightarrow \hskip .5cm
        \oint_c\sigma = \ \ \text{\emph{relative} invariant}
\end{equation}
Here the meaning of the r.h.s. of (\ref{jetotoiste4}) is as follows: take a cycle $c_1$
located in the hyper-plane $t=t_1$ and its image $c_2$ w.r.t. the flow of $\xi$
(it is located in the hyper-plane $t=t_2$).
Then integrals of $\sigma$ over $c_1$ and $c_2$ give the same number.
(Notice that $dt \wedge \hat \beta$ part of $\sigma$ does not contribute,
 since $dt$ vanishes on the hyper-planes.) So, indeed,
 statements (\ref{jetotoiste2}) and (\ref{jetotoiste4}) are,
 in this interpretation,
 equivalent.

First \emph{new} result by Cartan (w.r.t. Poincar\'e) is an observation
that more general interpretation of (\ref{jetotoiste4}) is possible.
Namely, take \emph{any} two cycles in $M \times \mathbb R$
which encircle common \emph{tube of solutions}
(here ``solutions'' mean integral curves of $\xi$,
 i.e. solutions of the dynamics as seen from $M \times \mathbb R$).
Then, \emph{still}, integrals of $\sigma$ over $c_1$ and $c_2$ give the same number.
See a proof in Appendix \ref{appproofintinv}.

Further Cartan's generalization, however, is much mo\-re interesting for us.
Recall that (\ref{definiciasigma1}) might also be regarded as a decomposition
of the\emph{ most general} $k$-form $\sigma$ on $M \times \mathbb R$,
see Appendix \ref{appdecomp}. In this case,
$\hat \alpha$ and $\hat \beta$ need not be obtained by pull-back from $M$.
Rather, they are the most general \emph{spatial} forms on $M \times \mathbb R$.
One easily sees that, in comparison with just pull-backs, they may be \emph{time-dependent},
i.e. it \emph{may happen} that
\begin{equation} \label{mayhappen}
        \mathcal L_{\partial_t}\hat \alpha  \neq 0
        \hskip 1cm
        \mathcal L_{\partial_t}\hat \beta  \neq 0
\end{equation}
(In coordinate presentation, their \emph{components} may depend on time.)

Recall that the proof of (\ref{jetotoiste4}) from Appendix \ref{appproofintinv}
did not use any details of the decomposition. The structure of the equation
(\ref{zakladnarovnica1}) is all one needs.
Notice, however, that the equivalence of
(\ref{zakladnarovnica1}) and (\ref{ivdalphajeexaktna}) is no longer true when
(\ref{mayhappen}) holds.
Instead, one easily computes (with the help of Appendix \ref{appdecomp})
that
\begin{equation} \label{isequivalent}
      i_\xi d\sigma = 0
      \hskip .7cm \Leftrightarrow \hskip .7cm
      {\mathcal L}_{\partial_t} \hat \alpha +i_v\hat d \hat \alpha  = \hat d\hat \beta
\end{equation}
(the term ${\mathcal L}_{\partial_t} \hat \alpha$ is new).
So, the equation
\begin{equation} \label{ixidesigma1}
      {\mathcal L}_{\partial_t} \hat \alpha +i_v\hat d \hat \alpha  = \hat d\hat \beta
\end{equation}
is \emph{the} equation that \emph{time-dependent} forms $\hat \alpha$ and $\hat \beta$
are to satisfy in order that integral of $\sigma$ is to be a relative integral invariant.

\subsection{\label{subsec:eulertimedependent}Non-stationary Euler equation}

Let us retell Cartan's results from the last section
in the context of hydrodynamics, i.e. for particular choice
(see Eq. (\ref{eulerstatform3}))
\begin{equation} \label{definiciasigmahydro}
       \sigma = \hat v - \mathcal E dt
\end{equation}
where, in usual coordinates $(\bold r,t)$ on $E^3\times \mathbb R$,
\begin{equation} \label{definiciahatv}
       \hat v := \bold v \cdot d\bold r \equiv \bold v (\bold r,t) \cdot d\bold r
\end{equation}
From (\ref{isequivalent}) we get
\begin{equation} \label{hydro1}
      i_\xi d\sigma = 0
      \hskip .7cm \Leftrightarrow \hskip .7cm
      {\mathcal L}_{\partial_t} \hat v +i_v\hat d \hat v  = - \hat d\mathcal E
\end{equation}
One easily checks (e.g. in Cartesian coordinates $(\bold r,t)$) that
\begin{equation} \label{hydro2}
      {\mathcal L}_{\partial_t} \hat v +i_v\hat d \hat v  = - \hat d\mathcal E
\end{equation}
is nothing but the complete, time-dependent, Euler equation (\ref{eulernorm}).
Therefore the time-dependent Euler equation may also be written in remarkably succinct form as
\begin{equation} \label{hydro3}
      i_\xi d\sigma = 0
      \hskip 1cm
      \text{\emph{Euler equation}}
\end{equation}
The form (\ref{hydro3}) of the Euler equation turns out to be very convenient
for analyzing some of its cosequences.
Two examples:

1. Just looking at (\ref{jetotoiste4}), (\ref{hydro3}) and (\ref{definiciasigmahydro}) one obtains
\begin{equation} \label{Kelvin_stat2}
                 \oint_c \bold v \cdot d\bold r = \text{const.}
                 \hskip 1.5cm \text{\emph{Kelvin's theorem}}
\end{equation}
(the two loops $c_1$ and $c_2$ are usually in constant-time hyper-planes $t=t_1$ and $t=t_2$).

2. Application of $d$ on both sides gives very quickly \emph{Helmholtz theorem}
(see the next Section \ref{subsec:helmholtztimedependent}).


\subsection{\label{subsec:helmholtztimedependent}Helmholtz statement on vortex lines - general case}

Application of $d$ on both sides of (\ref{hydro3}) and using formula
(\ref{cartanmagic}) results in
\begin{equation} \label{dsigmajeinvar1}
             \mathcal L_{\xi}(d\sigma) = 0
\end{equation}
This is, however, nothing but infinitesimal version of the statement
\begin{equation} \label{dsigmajeinvar2}
             \Phi_t^*(d\sigma) = d\sigma
             \hskip 2cm
             \Phi_t \leftrightarrow \xi
\end{equation}
or, in words, that the $d\sigma$ is \emph{invariant}
w.r.t. the flow of the fluid (regarded as the flow of $\xi$ on $M\times \mathbb R$).

Now, we want to see an integrable distribution behind vortex lines, again.
Define the distribution $\mathcal D$ in terms of annihilation of
as many as \emph{two} exact forms:
\begin{equation} \label{newdistributiondef}
         \mathcal D
         \hskip .4cm \leftrightarrow \hskip .4cm
         i_w d\sigma = 0 = i_w dt
\end{equation}
By repeating the reasoning from (\ref{identity1}) and (\ref{Disintegrable})
one concludes that $\mathcal D$ is \emph{integrable}.

The distribution $\mathcal D$ is, however, also \emph{invariant}
w.r.t. the flow of the fluid.
(Because of (\ref{dsigmajeinvar1}) and the trivial fact that $\mathcal L_{\xi}(dt) = 0$.)
So, integral submanifolds (surfaces) \emph{move with the fluid}.

What do they look like?
Although perhaps not visible at first sight,
they are nothing but vortex lines.

Indeed, making use of general formula (\ref{dongeneral}) from Appendix \ref{appdecomp}
and the form (\ref{hydro2}) of Euler equation we can write
\begin{eqnarray} 
      \label{dsigmavzdyvseob2}
      d\sigma &=& \hat d \hat v + dt \wedge ({\mathcal L}_{\partial_t} \hat v +\hat d \mathcal E)
                  \hskip .8cm \text{always} \\
  \label{dsigmanariesenivseob2}
        &=& \hat d \hat v +dt\wedge (-i_v\hat d\hat v)
                       \hskip 1.4cm \text{\emph{on solutions}}
\end{eqnarray}
Let us now contemplate Eq. (\ref{newdistributiondef}). It says, that the distribution consists
of \emph{spatial} vectors (i.e. those with vanishing \emph{time} component,
therefore annihilating $dt$) which, in addition, annihilate $d\sigma$.

Let $w$ be arbitrary \emph{spatial} vector.
Denote, for a while, $i_w\hat d \hat v =:\hat b$ (it is a \emph{spatial} 1-form).
Then, from (\ref{dsigmanariesenivseob2}),
\begin{equation} \label{iwdsigma}
  i_wd\sigma
  = \hat b -dt\wedge i_v\hat b
\end{equation}
from which immediately
\begin{equation} \label{iwdsigmaiszero}
             i_w(d\sigma) = 0
             \hskip 1cm \Leftrightarrow \hskip 1cm
             \hat b \equiv i_w\hat d \hat v = 0
\end{equation}
This says that we can, alternatively, describe the distribution $\mathcal D$
as consisting of those \emph{spatial} vectors which, in addition,
annihilate $\hat d \hat v$
(rather than $d\sigma$, as it is expressed in the definition (\ref{newdistributiondef})).
But Eqs. (\ref{definiciahatv}) and (\ref{vorticityform}) show that
\begin{equation} \label{vorticityform2}
             \hat d \hat v  = \boldsymbol \omega \cdot d\bold S
                       \equiv \boldsymbol \omega (\bold r,t) \cdot d\bold S
\end{equation}
so that $\hat d \hat v$ is nothing but the \emph{vorticity 2-form}
and, therefore, the integral surfaces of $\mathcal D$
may indeed be identified with vortex lines.
So, Helmholtz statement is also true in the general, time-dependent, case.
(Notice that the system of vortex lines looks, in general, different in different times.
This is because its generating object, the vorticity 2-form $\hat d \hat v$, depends on time.)

\subsection{\label{subsec:helmholtztubestimedependent}Helmholtz statement on vortex tubes - general case}

Vortex tube is a genuinely spatial concept and the statement concerns
purely kinematical property of \emph{any} velocity field at a single time
(see the beginning of Sec. \ref{subsec:helmholtztubestimeindependent}).
So, no (change of) dynamics has any influence on it.
If the statement were true before, it remains to be true now.

\section{\label{sec:surfaces}Generalization to surfaces}

In this section we present details concerning the surfaces mentioned in the Introduction.
By now it is easy, since we already know all the relevant ideas from hydrodynamics parts.

All symbols which occur here refer to objects mentioned in the
\emph{general theory of integral invariants} (due to Poincar\'e and Cartan, respectively,
i.e. objects from Sections \ref{subsec:poincare} and \ref{subsec:cartan})
rather than to their special instances used in hydrodynamics
(including the $n$-dimensional case).

\subsection{\label{subsec:surfacespoincare}Time-independent (Poincar\'e) case}

We apply $d$ on both sides of (\ref{ivdalphajeexaktna}) or (\ref{podmienka3})
and get
\begin{equation} \label{Lvdalphanula}
                 \mathcal L_v (d\alpha) = 0
\end{equation}
So, the $(k+1)$-form $d\alpha$ is invariant w.r.t. the flow generated on $M$ by $v$.

Now, define a distribution $\mathcal D$ given by annihilation of the form $d\alpha$:
\begin{equation} \label{distributiondalpha}
         \mathcal D
         :=
         \{ \text{vectors} \  w  \ \text{such that} \ \  i_w d\alpha = 0 \ \ \text{holds}  \}
\end{equation}
Its dimension is therefore
\begin{eqnarray} 
      \label{dimofDdalpha1}
          \text{dim} \ \mathcal D
      &=&  \text{dim} \ M - \text{rank} \ d\alpha     \\
  \label{dimofDdalpha2}
        &\le& \text{dim} \ M - (k+1)
\end{eqnarray}
(if $\alpha$ is $k$-form, see Appendix \ref{appdimensionfofD};
 the rank of $d\alpha$ is expected to be constant).

The distribution $\mathcal D$ has the following two properties.

First, it is \emph{invariant} w.r.t. the flow gene\-ra\-ted on $M$ by $v$.
(This is because of (\ref{Lvdalphanula}).)

Second, with the help of (\ref{identity1}) and (\ref{cartanmagic}) we see that
\begin{equation} \label{Dalphaisintegrable1}
           i_{w_1} d\alpha = 0 =
           i_{w_2} d\alpha
           \hskip .5cm \Rightarrow \hskip .5cm
         i_{[w_1,w_2]} d\alpha = 0
\end{equation}
i.e.
\begin{equation} \label{Dalphaisintegrable2}
           w_1, w_2 \in \mathcal D
           \hskip .5cm \Rightarrow \hskip .5cm
           [w_1, w_2] \in \mathcal D
\end{equation}
So, due to Frobenius criterion, $\mathcal D$ is \emph{integrable}.

Put the two properties together, this means that \emph{integral surfaces} (submanifolds)
of the distribution \emph{move with the} (abstract) \emph{fluid},
exactly in the spirit of the Helmholtz theorem on vortex lines.
(Notice that this behavior equally holds for any surface of \emph{smaller} dimension
 which resides within the maximal-dimension one.)

\begin{figure}[tb]
\begin{center}
\includegraphics[scale=0.30]{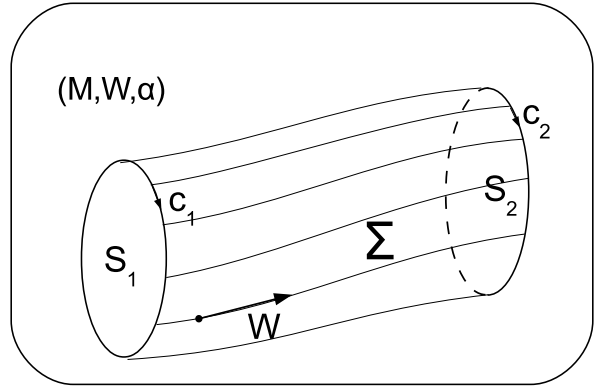}
\caption{Higher-dimensional analog of vortex tube, $\Sigma$.
         It is bounded by a $k$-dimensional boundary $c_1$ of a transversal $(k+1)$-dimensional surface $S_1$
         from the left and similarly by $c_2 \equiv \partial S_2$ from the right.
         Here $S_2 \equiv \Phi_s(S_1)$ for some $s$.}
\label{c1c2encircleSigma2}
\end{center}
\end{figure}
Now consider a vector \emph{field} $W\in \mathcal D$
(so it satisfies $i_Wd\alpha =0$;
 this is analog of the field $w$ directed along vortex lines,
 discussed in Sec. \ref{subsec:helmholtztubestimeindependent}).
Application of its flow $\Phi_s$ on $k$-dimensional boundary $c_1 \equiv \partial S_1$
of a transversal $(k+1)$-dimensional surface $S_1$ gives a $(k+1)$-dimensional analog of vortex tube, $\Sigma$
(see Fig.\ref{c1c2encircleSigma2}).
So
\begin{equation} \label{boundaryofV}
                 \partial \Sigma = c_1 - c_2
\end{equation}
Repeating either the reasoning from Appendix \ref{appproofintinv}
 or that from Sec. \ref{subsec:helmholtztubestimeindependent} we show that
 the (analog of the) \emph{strength} of the tube is constant along the tube
\begin{equation} \label{generaltubeHelmholtz}
                 \int_{S_1} d\alpha = \int_{S_2} d\alpha
\end{equation}
This is an analog of Helmholtz theo\-rem on vortex tubes.

\subsection{\label{subsec:surfacescartan}Time-dependent (Cartan) case}

We apply $d$ on both sides of (\ref{zakladnarovnica1}) and get
\begin{equation} \label{Lxidsigmanula}
                 \mathcal L_{\xi} (d\sigma) = 0
\end{equation}
So, the $(k+1)$-form $d\sigma$ is invariant w.r.t. the flow gene\-rated on $M\times \mathbb R$ by $\xi$.

Now, define a distribution $\mathcal D$ given by \emph{spatial} vectors
which annihilate the form $d\sigma$:
\begin{equation} \label{distributiondsigmadt}
         \mathcal D
         :=
         \{ \text{\emph{spatial} vectors} \  w  \ \text{such that} \ \
                                   i_w d\sigma = 0 \}
\end{equation}
Put another way, it is defined as
\begin{equation} \label{newdistributiondef2}
         w\in \mathcal D
         \hskip .4cm \Leftrightarrow \hskip .4cm
         i_w d\sigma = 0 = i_w dt
\end{equation}
The distribution is \emph{invariant} w.r.t. the flow gene\-ra\-ted on $M\times \mathbb R$ by $\xi$
(since \emph{both} its generating forms, $d\sigma$ as well as $dt$, are Lie-invariant
 w.r.t. $\xi$).

In addition, due to Frobenius criterion, the distribution is \emph{integrable}.
(One just applies (\ref{Dalphaisintegrable1}) to \emph{both} $d\sigma$ and $dt$.)

Put the two properties together, this means that integral submanifolds (surfaces)
of the distribution \emph{move with the} (abstract) \emph{fluid}
in the spirit of the Helmholtz theorem on vortex lines.

Finally, notice that, \emph{on solutions} of Eq. (\ref{zakladnarovnica1}),
the distribution $\mathcal D$ generated by the pair of forms $(d\sigma,dt)$
coincides with that generated by the pair $(\hat d\hat \alpha,dt)$.
(Just repeat argumentation in (\ref{dsigmavzdyvseob2}) - (\ref{iwdsigmaiszero})
 replacing $\hat v \mapsto \hat \alpha$, $\mathcal E \mapsto -\hat \beta$
 and Eq. (\ref{hydro2}) $\mapsto$ Eq. (\ref{ixidesigma1}).)
So it consists of \emph{spatial} vectors annihilating $\hat d \hat \alpha$.
Therefore, the statement about surfaces moving with the (abstract) fluid
here, in Sec. \ref{subsec:surfacescartan}, is a natural generalization
(namely to time-\emph{dependent} flow) of the corresponding statement mentioned
in Sec. \ref{subsec:surfacespoincare}.

Concerning the ``vortex tube'' Helmholtz theorem, it has nothing to do with dynamics
and therefore it is tri\-vially true also here (see Sec. \ref{subsec:helmholtztubestimedependent}).

\section{\label{sec:conclusions}Conclusions}

The main point discussed in this paper
is a statement concerning the \emph{general setting} of the
theory of \emph{integral invariants}
(rather than the
``ideal hydrodynamics on Riemanian manifolds'' \
or ``higher-dimensional hydrodynamics''
discussed, e.g., in Ref.~\onlinecite{ArnoldKhesin1998}
and in the numerous papers mentioned in references therein).

Namely, in the theory of integral invariants,
both the time-in\-de\-pendent version of Poincar\'e
and the extended, time-de\-pen\-dent version of Cartan, one can find specific
\emph{surfaces} which \emph{move with the} (abstract) ``\emph{fluid}''.

When the theory is applied to 3D-hydrodynamics of ideal and barotropic fluid
only subject to potential force, the surfaces become 1-dimensional and reduce to
well-known and useful concept of \emph{vortex lines}. Their property
of moving with the fluid (now the real one) becomes
the celebrated \emph{Helmholtz theorem} from 1858.

So, in this sense, the surfaces may be regarded as a generalization of the vortex lines.

One can also define, in the general higher-dimensional case,
an analog of the hydrodynamical concept of \emph{vortex tubes} and check that (an analog of)
Helmholtz theorem on \emph{strength} of the tubes is still true.

\begin{acknowledgments}
I acknowledge support from grant VEGA 1/0985/16.
\end{acknowledgments}

\appendix

\section{\label{appdecomp}Decomposition of forms}
\setcounter{equation}{0}

On $M\times \mathbb R$, a $p$-form $\alpha$ may be uniquely decomposed as
\begin{equation}
\alpha = dt \wedge \hat s + \hat r
\label{decomposition1}
\end{equation}
where both $\hat s$ and $\hat r$ are {\it spatial},
i.e. they do not contain the factor $dt$ in its coordinate presentation
(here, we assume \emph{adapted} coordinates, $t$ on $\mathbb R$ and some $x^i$ on $M$).
Simply, after writing the form in coordinates, one groups together all terms
 which do contain $dt$ once and, similarly, terms which do not contain $dt$ at all.
 Note, however, that $t$ still can enter {\it components} of any (even spatial) form.
 Therefore, when performing exterior derivative $d$ of a {\it spatial} form, say $\hat r$,
 there is a part, $\hat d \hat r$, which does not take into account the $t$-dependance
 of the components (if any; \emph{as if} it was performed \emph{just on} $M$), plus a part which,
 on the contrary, only operates on the $t$ variable. Putting both parts together, we have
 \begin{equation} \label{donspatial}
 d\hat r = dt \wedge \mathcal L_{\partial_t} \hat r + \hat d \hat r
\end{equation}
Then, for a general form (\ref{decomposition1}), we get
 \begin{equation} \label{dongeneral}
 d\alpha = dt \wedge (-\hat d \hat s + \mathcal L_{\partial_t} \hat r ) + \hat d \hat r
\end{equation}

\section{\label{appproofintinv}A proof of (\ref{jetotoiste4})}
\setcounter{equation}{0}

\begin{figure}[tb]
\begin{center}
\includegraphics[scale=0.30]{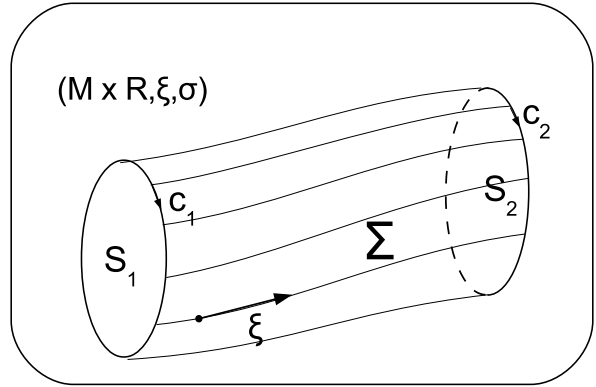}
\caption{$\Sigma$ is given by the family of trajectories (solutions) connecting $c_1$ and $c_2$,
         so that $\partial \Sigma = c_1 - c_2$. Integral of $d\sigma$ over $\Sigma$ vanishes,
         since $\xi$ is tangent to $\Sigma$ and annihilates $d\sigma$.}
\label{c1c2encircleSigma3}
\end{center}
\end{figure}

The proof is amazingly simple (see \$44 of Ref.~\onlinecite{Arnold1989}).
 Consider integral of $d\sigma$ over the $(k+1)$-chain $\Sigma$ given by the family of trajectories
 (solutions) connecting $c_1$ and $c_2$ (so that $\partial \Sigma = c_1 - c_2$,
  see Fig. \ref{c1c2encircleSigma3}.
Then
 $$
\begin{array} {rcl}
  \int_{\Sigma} d\sigma
       &\overset{1.} {=}& \int_{\partial \Sigma} \sigma = \oint_{c_1} \sigma - \oint_{c_2} \sigma \\
       &\overset{2.} {=}& 0
\end{array}
$$
The second line (zero) comes from observation, that $\xi$ is tangent to $\Sigma$ by construction,
so that integral of $d\sigma$ over $\Sigma$ consists of infinitesimal contributions proportional
to $d\sigma (\xi,\dots)$, all of them vanishing because of (\ref{zakladnarovnica1}).


\section{\label{appdimensionfofD}Dimension of the distribution $\mathcal D$}
\setcounter{equation}{0}

The distribution $\mathcal D$ from Eq. (\ref{distributiondalpha}) is given
as the \emph{kernel} of the linear map
 \begin{equation} \label{asthekernel}
   f: w \mapsto i_w d\alpha
\end{equation}
\emph{Rank} of the form $d\alpha$ is, by definition,
the dimension of the \emph{image} space of the map (\ref{asthekernel}).
Then, due to the standard rank-nullity theorem applied to $f$, we have
 \begin{equation} \label{ranknullity1}
   \text{dim} \ \text{Ker} \ f + \text{dim} \ \text{Im} \ f = \text{dim} \ T_xM \equiv \text{dim} \ M
\end{equation}
or
 \begin{equation} \label{ranknullity2}
   \text{dim} \ \mathcal D + \text{rank} \ d\alpha = \text{dim} \ M
\end{equation}
So, (\ref{dimofDdalpha1}) holds.

Now rank of a $p$-form is \emph{at least} $p$
(it is $p$ for \emph{decomposable} form), so
 \begin{equation} \label{rankofdalpha}
   \text{rank} \ d\alpha \ge k+1
\end{equation}
and, therefore, (\ref{dimofDdalpha2}) holds.

The dimension of $\mathcal D$ in time-dependent case
(i.e. given by (\ref{distributiondsigmadt}) or (\ref{newdistributiondef2}))
equals the dimension of $\mathcal D$ from the time-independent case
(given by (\ref{distributiondalpha})).
Indeed, as is mentioned in the last paragraph of Sec. \ref{subsec:surfacescartan},
the distribution $\mathcal D$ on $M\times \mathbb R$ generated by the pair of forms $(d\sigma,dt)$
coincides, on solutions, with that generated by the pair $(\hat d\hat \alpha,dt)$.
So it consists of \emph{spatial} vectors annihilating $\hat d \hat \alpha$.
Or, when thinking of \emph{dimensions} alone,
of vectors on $M$ annihilating $d \alpha$.

\nocite{*}
\bibliography{generalized_vortex_for_JGP_2a}

\end{document}